\documentstyle[prd,aps,preprint,tighten]{revtex}
\begin{document}
\preprint{                                                BARI-TH/291-97}
\draft
\title{	Zenith distribution of atmospheric neutrino events \\
		and electron neutrino mixing }
\author{    G.~L.~Fogli, E.~Lisi, A.~Marrone, and D.~Montanino          }
\address{   Dipartimento di Fisica and Sezione INFN di Bari, 		\\
                  Via Amendola 173, I-70126 Bari, Italy			}
\maketitle
\begin{abstract}
Assuming atmospheric neutrino oscillations with dominant 
$\nu_\mu\leftrightarrow\nu_\tau$ transitions, we discuss how subdominant
$\nu_e$ mixing (within the Chooz reactor bounds) can alter the zenith 
distributions of neutrino-induced electrons and muons. We isolate two 
peculiar distortion effects, one mainly related to $\nu_e$ mixing in 
vacuum and the other to matter oscillations, that may be sufficiently 
large to be detected by the SuperKamiokande atmospheric $\nu$ experiment. 
These effects (absent for pure two-flavor $\nu_\mu\leftrightarrow\nu_\tau$ 
transitions) do not vanish in the limit of energy-averaged oscillations.
\end{abstract}

	The muon-electron flavor anomaly of atmospheric neutrino events, and 
its possible dependence on the zenith angle $\Theta$ (i.e., on the neutrino 
pathlength $L$) represent tentative evidence for 
$\nu_\mu\leftrightarrow\nu_{e,\mu,\tau}$ oscillations with neutrinos 
square mass differences $m^2 \gtrsim 10^{-3}$ eV$^2$ and large mixing
$\sin^2 2\theta$ (see, e.g., \cite{Fo97} and references therein). 
The recent negative results from the Chooz reactor experiment \cite{Ch97}
in the $\nu_e\leftrightarrow\nu_e$ disappearance channel,
\begin{equation}
\sin^2 2\theta_{ee} \lesssim 0.18
{\rm\ \ at\ 90\%\  C.L.\ for\ \ } m^2\gtrsim
10^{-3}\ {\rm eV}^2\ ,  
\label{chooz}
\end{equation}
exclude $\nu_\mu\leftrightarrow\nu_e$ oscillations as a viable explanation
of the atmospheric anomaly. However, scenarios dominated by 
$\nu_\mu\leftrightarrow\nu_\tau$ oscillations in the
range $m^2\gtrsim 10^{-3}$ eV$^2$ are still practicable, and 
provide good fits to the preliminary SuperKamiokande data 
for $\sin^2 2\theta_{\mu\tau} \sim 1$\cite{To97}.

	In this paper we show how subdominant $\nu_e$ mixing can induce 
nonnegligible effects on the zenith distributions of atmospheric neutrino 
events.  In particular, we discuss two effects that, interestingly,  do not 
vanish  in the limit of energy-averaged oscillations. The first is
mainly related to $\nu_e$ mixing in vacuum and the second represents a 
genuine three-flavor matter effect.  Although these effects are implicitly
taken into account in detailed numerical calculations of three-flavor neutrino
oscillations \cite{Fo97}, we think it useful to describe them separately,
since their dependence on the neutrino energy $E$ or pathlength
$L$, which is very different from the usual (vacuum) $L/E$ form, could
play a role in the interpretation of recent, high
statistics atmospheric neutrino data.

	We adopt, as in previous works \cite{Fo97,Fo94},  a three-flavor 
framework with one dominant mass scale, i.e., a neutrino mass spectrum 
with two light states $(\nu_1,\,\nu_2)$ and one heavy state $\nu_3$:
\begin{equation}
|m^2_2-m^2_1| \ll |m^2_3 - m^2_{1,2}| = m^2 \gtrsim 10^{-3} {\rm\ 
\ eV}^2\ .
\end{equation}
In addition, we assume that the mixing matrix element $U_{e3}$ is small
and that $\nu_\mu$ and $\nu_\tau$ are maximally mixed,
\begin{equation} 
U^2_{e3} = \epsilon\,,\; U^2_{\mu3}=U^2_{\tau3}=(1-\epsilon)/2\ ,
\end{equation}
where the free parameter $\epsilon$ is constrained by Eq.~(\ref{chooz})
in the range
\begin{equation}
\epsilon \lesssim  0.05\ ,
\label{eps}
\end{equation}
the mixing angle probed by Chooz being
$\sin^2 2\theta_{ee}=4\,U^2_{e3}\,(1-U^2_{e3})=4\,\epsilon\,(1-\epsilon)$.
In the following, we shall neglect terms of the order $\epsilon^2$.
Notice that
$\sin^2 2\theta_{\mu\tau}=4 U^2_{\mu3}U^2_{\tau3}\simeq 
1-2\epsilon \gtrsim 0.9$,
consistently with preliminary fits to the SuperKamiokande data \cite{To97}.
(In particular, the case $\epsilon=0$ corresponds to pure two-flavor 
$\nu_\mu\leftrightarrow\nu_\tau$ oscillations with maximal mixing.)

	In this framework, the flavor oscillation probabilities 
in vacuum $P_{\alpha\beta}^{\rm vac}$ assume the simple form
\begin{equation}
\left(\begin{array}{ccc}
P_{ee} & P_{e\mu} & P_{e\tau} \\
P_{\mu e} & P_{\mu\mu} & P_{\mu\tau} \\
P_{\tau e} & P_{\tau\mu} & P_{\tau\tau} \\
\end{array}
\right)_{\rm vac}
=\left(\begin{array}{ccc}
1-4\epsilon S &2\epsilon S& 2\epsilon S\\
2\epsilon S & 1- S & (1-2\epsilon) S\\
2\epsilon S &(1-2\epsilon) S & 1- S
\end{array}
\right)\ ,
\label{pvac}
\end{equation}
where $S$ is the usual oscillation factor $S=\sin^2(1.27 m^2 L/E)$
and $m^2$, $L$, and $E$ are measured in eV$^2$, km, and GeV, respectively.

	We remind that, in the presence of oscillations, the ratios 
$R_{e,\mu}$ of neutrino-induced lepton event rates with oscillations 
$(N_{e,\mu})$ and without oscillations
$(N^0_{e,\mu})$ are approximately given by (see, e.g., \cite{Fo95})
\begin{eqnarray}
\label{re}
R_e &=& N_e/N^0_e \simeq P_{ee} + r P_{\mu e}\ ,\\
R_\mu &=& N_\mu/N^0_\mu \simeq P_{\mu\mu} + P_{e\mu}/r\ ,
\label{rmu}
\end{eqnarray}
where $r\sim N^0_\mu/N^0_e$.
In order to isolate the effects that we are going to
discuss, we work in the regime of energy-averaged vacuum oscillations,
i.e., $S\simeq 1/2$ (removal of this approximation will be discussed at the
end). In this case the averaged vacuum probabilities are 
given by
\begin{equation}
\overline P_{\alpha \beta}^{\rm vac} \simeq
\left(\begin{array}{ccc}
1-2\epsilon  & \epsilon & \epsilon \\
\epsilon  & 1/2 & 1/2-\epsilon\\
\epsilon  &1/2-\epsilon  & 1/2
\end{array}
\right)\ .
\label{pave}
\end{equation}

	The first effect that we discuss is related to the zenith dependence
of the parameter $r$. For subGeV events, it is $r\sim 2$ without appreciable
zenith dependence \cite{Ag96}. 
However, for energies in the multiGeV (or higher)
range, the  ratio  $r\sim N^0_\mu/N^0_e$ does depend
on the zenith angle $\Theta$ and, in particular, it increases when
going from the horizontal $(H)$ direction, $\cos\Theta\simeq0$,
to the vertical $(V)$ up and down directions, $\cos\Theta\simeq\pm 1$:
\begin{equation}
r_H < r_V\ .
\end{equation}
In fact, the atmospheric $\nu_e$ and $\nu_\mu$ fluxes both decrease
towards the vertical (where the slanted depth in the atmosphere is reduced),
but $\nu_e$'s are more effectively  suppressed than $\nu_\mu$'s due to their
different parent decay chains. Moreover,
the greater the energy of the parents, 
the longer the decay lengths, the stronger 
the dependence of $r$ on the slanted depth and thus on the zenith angle
$\Theta$.

	Therefore, even for averaged ($\Theta$-independent) vacuum 
oscillations, the quantities $R_e$ and $R_\mu$ in Eqs~(\ref{re},\ref{rmu})
acquire a dependence on $\Theta$ through the parameter $r=r(\Theta)$. 
For instance, in the multiGeV energy range one has approximately 
(using the angular $\nu$ spectra of \cite{Ag96})
\begin{equation}
	r_H \sim 2\,,\; r_V\sim 4\ ,
\end{equation}
which, using Eqs~(\ref{re},\ref{rmu},\ref{pave}), imply
a vertical-horizontal asymmetry $A_{\rm VH}$ of the normalized
rates $R_{e,\mu}$ that would be
absent for pure two-flavor $\nu_\mu\leftrightarrow\nu_\tau$
oscillations $(\epsilon=0)$:
\begin{eqnarray}
A^e_{\rm VH} &=& R_e^V/R_e^H-1\simeq \epsilon (r_V-r_H)
\simeq 2\epsilon\ ,\\
A^\mu_{\rm VH} &=& R_\mu^V/R_\mu^H-1\simeq -2\epsilon (r_H^{-1}-r_V^{-1})
\simeq -\epsilon/2\ .
\end{eqnarray}
Given the constraint of Eq.~(\ref{eps}), the asymmetry 
can be as large as +10\% ($-2.5$\%) for electrons (muons) and thus it
might be detected in SuperKamiokande.

	The second effect is a peculiar three-flavor oscillation effect
in matter. It has been described at the end of Appendix C in Ref.~\cite{Fo97} 
and also recently discussed in Ref.~\cite{Gi97}, building upon an
earlier suggestion by J.\ Pantaleone \cite{Pa94}.  It is based 
on the fact that the degenerate
doublet of light neutrinos $(\nu_1,\,\nu_2)$ splits in matter, and the
oscillating term driven by the 1--2 effective mass splitting becomes 
energy-independent in the limit $m^2 \gg  A=2\sqrt{2} G_F N_e E$,
where $N_e$ is the electron density in the Earth. In this limit,
and for constant $N_e$, the difference between the
oscillation probability in matter $\overline P_{\alpha\beta}^{\rm mat}$
and in vacuum  $\overline P_{\alpha\beta}^{\rm vac}$
is given by a simple expression (see Appendix C of \cite{Fo97}),
\begin{equation}
\left(\begin{array}{ccc}
\overline P_{ee} & \overline P_{e\mu} &\overline  P_{e\tau} \\
\overline P_{\mu e} &\overline  P_{\mu\mu} &\overline  P_{\mu\tau} \\
\overline P_{\tau e} & \overline P_{\tau\mu} & \overline P_{\tau\tau} \\
\end{array}
\right)_{\rm mat}=
\left(\begin{array}{ccc}
\overline P_{ee} & \overline P_{e\mu} & \overline P_{e\tau} \\
\overline P_{\mu e} & \overline P_{\mu\mu} & \overline P_{\mu\tau} \\
\overline P_{\tau e} & \overline P_{\tau\mu} & \overline P_{\tau\tau} \\
\end{array}
\right)_{\rm vac}
+
\left(\begin{array}{ccc}
0 & 0 & 0 \\
0 & -\delta P & +\delta P \\
0 & +\delta P & -\delta P \\
\end{array}
\right)
\ ,
\label{pmat}
\end{equation}
where 
\begin{eqnarray}
\delta P&=&4\,\frac{U^2_{e3}\, U^2_{\mu3}\, U^2_{\tau3}}{(1-U^2_{e3})^2}\sin^2
\left(
2^{-\frac{1}{2}}\, G_F\, N_e\, (1-U^2_{e3}) L
\right)\\
&\simeq& \epsilon\, \sin^2\left(
2.47\, (1-\epsilon)\, \frac{N_e}{{\rm mol/cm}^3}\,\cos\Theta
\right)\ ,
\label{deltap}
\end{eqnarray}
by taking approximately $L\simeq 2 R_{\oplus} \cos\Theta$.

	We stress that this matter effect is basically different from the
``usual'' resonance effect, which is driven by the largest mass splitting
and occurs when $m^2 \sim A$. The correction $\delta P$, instead, is
induced by the two light mass states even if they are exactly degenerate
in vacuum. Moreover, $\delta P$ has the same form for both neutrinos
and antineutrinos \cite{Fo97}, while the usual enhancement of oscillations 
in matter cannot occur for both $\nu$'s and $\overline\nu$'s. Also notice that
$\delta P$ vanishes in the limit of pure $2\nu$ oscillations,
i.e., when one of the $U_{\alpha3}$ is zero.

	The consequences of the additional term $\delta P$ on the lepton 
spectra may be rather interesting. In fact,
the argument of the $\sin^2$ term in Eq.~(\ref{deltap}) is
large for typical electron densities in the Earth ($\sim$2--6 mol/cm$^3$).
Moreover, it is energy-independent and thus is not smeared by the
broad neutrino energy spectrum. Since the phase of $\delta P$
depends only on the neutrino
direction, its effects should be revealed more easily in the angular
spectrum of upward-going muons, where the $\mu$ and $\nu_\mu$ directions
are highly correlated. (Notice that electron spectra are unaffected by
$\delta P$.)  However, the amplitude of this 
``geometrical'' oscillation is small  ($\lesssim \epsilon$), and it
might be difficult to unfold it from $\Theta$ distributions. 
It should be easier to investigate the global effects of $\delta P$
on integral quantities, such as the up-down asymmetry of contained events
recently proposed in \cite{Fl97}.

	In particular, the up-going $(U)$ muon rate differs from the 
down-going $D$ muon rate through the term $\delta P$,
\begin{eqnarray}
R^D_\mu &=& \overline P_{\mu\mu}^{\rm vac}+
\overline P_{e\mu}^{\rm vac}/r\ ,\\
R^U_\mu &=& \overline P_{\mu\mu}^{\rm vac}+
\overline P_{e\mu}^{\rm vac}/r -\delta P\ ,
\end{eqnarray}
so that, defining $U_\mu$ and $D_\mu$ as
 the integral muon rates in the two hemispheres below and above the horizon 
$(U_\mu=\int\! d\Omega_U\, R_\mu^U,\, 
D_\mu=\int\! d\Omega_D\, R_\mu^D)$, one gets a nozero up-down muon
asymmetry $A^\mu_{\rm UD}$:
\begin{equation}
A^\mu_{\rm UD} = \frac{U_\mu}{D_\mu}-1=\frac{-\int\!d\Omega_U\, \delta P}
{2\pi(\overline P_{\mu\mu}^{\rm vac}+\overline P_{e\mu}^{\rm vac}/r)}\simeq
-\epsilon
\end{equation}
(up to a negligible term depending on $N_e$)
which, given the bound in Eq.~(\ref{eps}), can be as large as
a few percent and thus possibly detectable in the SuperKamiokande
experiment.

	In conclusion, we remind that the approximation of energy-averaged
oscillations (or, equivalently, of large $m^2$) has been used only to
isolate more clearly  the two effects discussed in this work. For values of
$m^2$ in the range suggested by the SuperKamiokande preliminary fits 
($m^2\sim10^{-3}$--$10^{-2}$ eV$^2$) \cite{To97}, these 
peculiar effects are generally entangled with the ``ordinary'' vacuum 
(and possibly matter)
oscillation effects, and it may be difficult to separate each of them, 
although the different functional dependences on $E$ or $L$ should help. 
The main message of this work is that, even if vacuum oscillations in the
$\nu_\mu\leftrightarrow\nu_\tau$ channel (with the usual
$L/E$ dependence) are expected to
give a dominant contribution to the atmospheric neutrino
anomalies, a small (few percent)  $\nu_e$ admixture can produce
additional, interesting effects that can alter the simple $L/E$ dependence 
of the lepton spectra at a level detectable in SuperKamiokande, 
without violating the recent limits placed by the
Chooz reactor experiment. We plan to investigate both dominant 
and subdominant oscillation effects in SuperKamiokande through more 
complete numerical calculations, when a more detailed description
of this experiment and of its results will become available.

We thank A.\ Yu.\ Smirnov for careful reading of the manuscript.
One of us (EL) thanks the Abdus Salam International
Centre for Theoretical Physics
(ICTP, Trieste) for kind hospitality during the preparation of this
work, and T.\ Montaruli, S.\ Petcov, and A.\ Yu.\ Smirnov for helpful
discussions.


\end{document}